\documentclass[aps,twocolumn,amssymb,amsmath,superscriptaddress]{revtex4-1}

\usepackage{bm}
\usepackage{color}
\usepackage{graphicx}
\usepackage{dcolumn}
\usepackage{graphicx}
\usepackage[colorlinks=true, letterpaper=true, pdfstartview=FitV, linkcolor=blue, citecolor=blue, urlcolor=blue]{hyperref}
\usepackage[normalem]{ulem}
\usepackage{amsmath}
\usepackage{epstopdf}
\usepackage{multirow}

\setcitestyle{square}

\makeatletter
\renewcommand\@biblabel[1]{#1.}
\makeatother

\begin{document}

\title{Strong magneto-optical and anomalous transport manifestations in two-dimensional van der Waals magnets Fe$_n$GeTe$_2$ ($n$ = 3, 4, 5)}

\author{Xiuxian Yang}
\affiliation{Centre for Quantum Physics, Key Laboratory of Advanced Optoelectronic Quantum Architecture and Measurement (MOE), School of Physics, Beijing Institute of Technology, Beijing, 100081, China}
\affiliation{Beijing Key Lab of Nanophotonics \& Ultrafine Optoelectronic Systems, School of Physics, Beijing Institute of Technology, Beijing, 100081, China}
\affiliation{Kunming Institute of Physics, Kunming 650223, China}

\author{Xiaodong Zhou}
\affiliation{Centre for Quantum Physics, Key Laboratory of Advanced Optoelectronic Quantum Architecture and Measurement (MOE), School of Physics, Beijing Institute of Technology, Beijing, 100081, China}
\affiliation{Beijing Key Lab of Nanophotonics \& Ultrafine Optoelectronic Systems, School of Physics, Beijing Institute of Technology, Beijing, 100081, China}

\author{Wanxiang Feng}
\email{wxfeng@bit.edu.cn}
\affiliation{Centre for Quantum Physics, Key Laboratory of Advanced Optoelectronic Quantum Architecture and Measurement (MOE), School of Physics, Beijing Institute of Technology, Beijing, 100081, China}
\affiliation{Beijing Key Lab of Nanophotonics \& Ultrafine Optoelectronic Systems, School of Physics, Beijing Institute of Technology, Beijing, 100081, China}

\author{Yugui Yao}
\affiliation{Centre for Quantum Physics, Key Laboratory of Advanced Optoelectronic Quantum Architecture and Measurement (MOE), School of Physics, Beijing Institute of Technology, Beijing, 100081, China}
\affiliation{Beijing Key Lab of Nanophotonics \& Ultrafine Optoelectronic Systems, School of Physics, Beijing Institute of Technology, Beijing, 100081, China}

\date{\today}

\begin{abstract}
Two-dimensional ferromagnetic materials have recently been attracted much attention mainly due to their promising applications towards multi-functional devices, e.g., microelectronics, spintronics, and thermoelectric devices.  Utilizing the first-principles calculations together with the group theory analysis, we systematically investigate the magnetocrystalline anisotropy energy, magneto-optical effect, and anomalous transport properties (including anomalous Hall, Nernst, and thermal Hall effects) of monolayer and bilayer Fe$_n$GeTe$_2$ ($n$ = 3, 4, 5).  The monolayer Fe$_n$GeTe$_2$ ($n$ = 3, 4, 5) exhibits the out-of-plane, in-plane, and in-plane ferromagnetic orders with considerable magnetocrystalline anisotropy energies of -3.17, 4.42, and 0.58 meV/f.u., respectively.  Ferromagnetic order is predicted in bilayer Fe$_4$GeTe$_2$ while antiferromagnetic order prefers in bilayer Fe$_3$GeTe$_2$ and Fe$_5$GeTe$_2$.  The group theory analysis reveals that in addition to monolayer ferromagnetic Fe$_n$GeTe$_2$ ($n$ = 3, 4, 5), the magneto-optical and anomalous transport phenomena surprisingly exist in bilayer antiferromagnetic Fe$_5$GeTe$_2$, which is much rare in realistic collinear antiferromagnets.  If spin magnetic moments of monolayer and bilayer Fe$_n$GeTe$_2$ are reoriented from the in-plane to out-of-plane direction, the magneto-optical and anomalous transport properties enhance significantly, presenting strong magnetic anisotropy.  We also demonstrate that the anomalous Hall effect decreases with the temperature increases.  The gigantic anomalous Nernst and thermal Hall effects are found in monolayer and bilayer ferromagnetic Fe$_n$GeTe$_2$, and the largest anomalous Nernst and thermal Hall conductivities, respectively, of -3.31 A/Km and 0.22 W/Km at 130 K are observed in bilayer ferromagnetic Fe$_4$GeTe$_2$.  Especially, bilayer antiferromagnetic Fe$_5$GeTe$_2$ exhibits large zero-temperature anomalous Hall conductivity of 2.63 e$^2$/h as well as anomalous Nernst and thermal Hall conductivities of 2.76 A/Km and 0.10 W/Km at 130 K, respectively.  Our results suggest that two-dimensional van der Waals magnets Fe$_n$GeTe$_2$ ($n$ = 3, 4, 5) have great potential applications in magneto-optical devices, spintronics, and spin caloritronics.
\end{abstract}

\maketitle

\section{Introduction}\label{intro}
According to the Mermin-Wagner theorem, ferromagnetism is hardly survived in two-dimensional (2D) system due to the enhanced thermal fluctuations~\cite{Mermin1966}.  Recently, exfoliating 2D magnetic materials from van der Waals (vdW) layered magnets has been confirmed to be an efficient method.  For example, the 2D materials, CrI$_3$~\cite{Huang2017} and Cr$_2$Ge$_2$Te$_6$~\cite{Gong2017}, were first experimentally reported to have the long-range magnetic order at low temperature in 2017.  The existence of atomically thin magnetic materials reveals that the magnetic anisotropy could counter thermal fluctuations and thus stabilized the long-range magnetic order.  Therefore, the 2D magnetic materials have attracted huge interest and are expected to be used in various multi-functional devices, such as microelectronics, spintronics, and thermoelectric devices.  However, due to the low Curie temperature ($T_C$) of CrI$_3$ ($\sim$ 45 K)~\cite{Huang2017} and Cr$_2$Ge$_2$Te$_6$ ($\sim$ 65 K)~\cite{Gong2017}, it poses a big challenge for using 2D vdW magnetic materials in any realistic applications.  In 2016, through the density-functional theory calculations, the monolayer (ML) Fe$_3$GeTe$_2$ (F3GT) was predicted to be a stable 2D magnetic system with large magnetocrystalline anisotropy energy (MAE)~\cite{Zhuang2016}, which provides a theory basis for experimental exfoliation.  Subsequently, the 2D vdW layered material, F3GT, was successfully exfoliated from the bulk materials and the $T_C$ of exfoliated ML is 130 K~\cite{Fei2018}.  Additionally, the $T_C$ of trilayer F3GT can be imporved to the room temperature by ionic-liquid gating technology~\cite{Deng2018}. As the first 2D ferromagnetic metal, F3GT has been extensively studied, such as the electronic-, optical-, transport-, and spin-related properties~\cite{Zhuang2016,Johansen2019,Wang2019,Albarakati2019,Lin2019,JiangMC2020,Xu2019,LiYin2020,Zheng2020}.  

The exploration of 2D vdW magnetic materials with higher $T_C$ has never stopped.  Following F3GT, ferromagnetic Fe$_{5-x}$GeTe$_2$ (sometimes referred to as Fe$_5$GeTe$_2$, F5GT) and Fe$_4$GeTe$_2$ (F4GT) have also been synthesized~\cite{Stahl2018,May2019,Seo2020}.  By reflective magnetic circular dichroism and Hall effect measurements, F5GT exhibits intrinsic ferromagnetic order at room temperature in bulk material ($T_C$ = 310 K) and nanoflakes ($T_C$ = 270 $\sim$ 300 K), respectively~\cite{May2019}.  Similarly, F4GT also shows high $T_C$ (270 K) in the thin flakes~\cite{Seo2020}.  Compared with F3GT, F4GT and F5GT have higher $T_C$, and therefore they may be more suitable for realistic device applications.  However, with the increase of iron occupancy, the magnetic properties of F4GT and F5GT turn to be more complex.  For example, in nanoflakes F4GT (thick than seven-layer), the spin changes from the out-of-plane to in-plane direction around 110 K, while the spin-reorientation temperature is significantly enhanced up to 200 K if the thickness of nanoflakes reduces down to seven-layer~\cite{Seo2020}.  The out-of-plane magnetization also emerges in the nanoflakes F5GT and magnetic anisotropy is enhanced with decreasing thickness down to five unit-cell layer~\cite{Ohta_2020}.  Besides the temperature- and thickness-dependent magnetic anisotropy, other fundamental physical manifestations, such as magneto-optical response as well as anomalous electronic, thermal, and thermoelectric transports, are still unclear in the atomically thin limit of Fe$_n$GeTe$_2$ ($n$ = 3, 4, 5), especially for F4GT and F5GT.  It essentially hinders the practical applications of Fe$_n$GeTe$_2$, and therefore it is time to uncover these physical properties of Fe$_n$GeTe$_2$.

In this paper, utilizing the first-principles calculations, we first determine the magnetic ground states of ML and bilayer (BL) Fe$_n$GeTe$_2$ ($n$ = 3, 4, 5).  The calculated MAE indicates that ML F3GT exhibit strong out-of-plane ferromagnetic order, while ML F4GT and F5GT show in-plane ferromagnetic order.  The out-of-plane magnetization of Fe$_n$GeTe$_2$ ($n$ = 3, 4, 5) increases with increasing the number of layers.  For BL structures, F3GT and F5GT exhibit interlayer antiferromagnetic coupling, while F4GT shows strong interlayer ferromagnetic coupling.  Subsequently, we use magnetic group theory to analyze the shape of optical conductivity tensor, and hence reveal the symmetry requirements for magneto-optical effect (MOE) and anomalous magnetotransport phenomena, including anomalous Hall effect (AHE), anomalous Nernst effect (ANE), and anomalous thermal Hall effect (ATHE).  These effects are symmetrically allowed in ML ferromagnetic F3GT with out-of-plane magnetization, and in ML ferromagnetic Fe$_n$GeTe$_2$ ($n$ = 4, 5) and BL ferromagnetic Fe$_n$GeTe$_2$ ($n$ = 3, 4, 5) with both in-plane and out-of-plane magnetization.  More intriguing, we find that BL antiferromagnetic F5GT satisfies the symmetry requirements and can appear MOE, AHE, ANE, and ATHE, whereas these effects are prohibited in BL antiferromagnetic F3GT and F4GT.  For the magnetic structures that satisfy the symmetry requirements, we calculate the magneto-optical Kerr and Faraday spectra, zero- and finite-temperature anomalous Hall conductivity (AHC), anomalous Nernst conductivity (ANC), and anomalous thermal Hall conductivity (ATHC).  The dependence of MOE, AHE, ANE, and ATHE on magnetization direction and thin-film thickness are clearly observed in Fe$_n$GeTe$_2$.  The calculated magneto-optical and anomalous transport properties are impressively large, which are comparable to or even larger than that of many famous ferromagnets and antiferromagnets. Our results suggest that 2D vdW magnets Fe$_n$GeTe$_2$ have potential applications in magneto-optical devices, spintronics, and spin caloritronics.

\section{Theory and computational details}\label{method}

The MOE is one of the fundamental physical effects in condensed-matter physics, which can be described as that when a linearly polarized light hits on the surface of a magnetic material, the polarization planes of reflected and transmitted lights shall rotate, called magneto-optical Kerr~\cite{Kerr1877} and Faraday~\cite{Faraday1846} effects, respectively. Nowadays, the MOE has become to be a powerful and nondestructive probe of magnetism in 2D materials~\cite{Huang2017,Gong2017}. To quantitatively characterize the magneto-optical performance of a given magnetic material, the Kerr ($\theta_\textnormal{K}$) and Faraday ($\theta_\textnormal{F}$) rotation angles, the deflection of the polarization planes with respect to the incident light, are used.  Moreover, the $\theta_\textnormal{K(F)}$ (rotation angle) and $\varepsilon_\textnormal{K(F)}$ (ellipticity) are usually combined into the so-called complex Kerr (Faraday) angle, written as~\cite{Suzuki1992,Guo1995,Ravindran1999}
\begin{equation}\label{eq:Kerr}
\phi_\textnormal{K}=\theta_\textnormal{K}+i\varepsilon_\textnormal{K}=i\frac{2\omega d}{c}\frac{\sigma_{xy}}{\sigma_{xx}^s},
\end{equation}
and 
\begin{equation}\label{eq:Faraday}
\phi_\textnormal{F}=\theta_\textnormal{F}+i\varepsilon_\textnormal{F}=\frac{\omega d}{2c}(n_+-n_-), 
\end{equation}
\begin{equation}\label{eq:Faraday2}
n_\pm^2=1+\frac{4\pi i}{\omega}(\sigma_{xx}\pm i\sigma_{xy}),
\end{equation}
where $c$, $\omega$, and $d$ are the speed of light in vacuum, frequency of incident light, and thin-film thickness, respectively.  $\sigma_{xy}$ is the off-diagonal element of the optical conductivity tensor for a magnetic thin-film and $\sigma_{xx}^s$ is the diagonal element of the optical conductivity tensor for a nonmagnetic substrate (e.g., SiO$_2$).  SiO$_2$ is a large band gap insulator and $\sigma_{xx}^s = i(1-n^2)\omega/4\pi$ with the refractive index of $n$ = 1.546.   $n_\pm^2$ are the eigenvalues of the dielectric tensor for a magnetic thin-film.  From the expressions of complex Kerr and Faraday angles [Eq.~\eqref{eq:Kerr}--\eqref{eq:Faraday2}], one can see that the crucial quantity is the optical conductivity, which can be calculated by Kubo-Greenwood formula~\cite{Arash2008,Yates2007}, 
\begin{eqnarray}\label{eq:OPC}
\sigma_{\alpha\beta}&=&\sigma_{\alpha\beta}^1 (\omega) + i\sigma_{\alpha\beta}^2 (\omega) \nonumber\\
&=& \frac{ie^2\hbar}{N_k\Omega_c}\sum_{\textbf{k}}\sum_{n, m}\frac{f_{m\textbf{k}}-f_{n\textbf{k}}}{\varepsilon_{m\textbf{k}}-\varepsilon_{n\textbf{k}}} \nonumber\\
&&\times\frac{\langle\psi_{n\textbf{k}}|\hat{v}_{\alpha}|\psi_{m\textbf{k}}\rangle\langle\psi_{m\textbf{k}}|\hat{v}_{\beta}|\psi_{n\textbf{k}}\rangle}{\varepsilon_{m\textbf{k}}-\varepsilon_{n\textbf{k}}-(\hbar\omega+i\eta)},
\end{eqnarray}
where the subscripts $\alpha,\beta\in \lbrace x,y\rbrace$.  The superscripts $1$ and $2$ represent the real and imaginary parts of $\sigma_{xy}$, $\Omega_c$ is the volume of unit cell and $N_k$ is the total number of $k$-points for sampling the Brillouin zone (here a $k$-mesh of 300$\times$300$\times$1 is used),  $\psi_{n\textbf{k}}$ and $\varepsilon_{n\textbf{k}}$ are the Wannier function and interpolated energy at the band index $n$ and momentum $\textbf{k}$, respectively,  $\hat{v}_{x,y}$ are the velocity operators along the $x$ or $y$ direction, $\hbar\omega$ is the photon energy, $\eta$ is an adjustable parameter for energy smearing (here $\eta$ = 0.15 eV), and $f_{n\textbf{k}}$ is the Fermi-Dirac distribution function.

In condensed-matter physics, the AHE is another important physical phenomenon, which is characterized by a transverse voltage drop generated by a longitudinal charge current in the absence of an external magnetic field~\cite{Nagaosa2010}.  Physically speaking, the dc limit of the real part of the off-diagonal element of optical conductivity, $\sigma_{xy}^1(\omega \rightarrow 0)$, is nothing but the intrinsic AHC.  Therefore, the AHE and MOE are closely related to each other and have similar physical origins.  Based on the linear response theory, the intrinsic AHC at zero-temperature ($\sigma^\textnormal{T=0}_{xy}$) can be calculated~\cite{Yao2004},
\begin{equation}\label{eq:AHC}
\sigma^\textnormal{T=0}_{xy}=-\frac{e^2}{\hbar V}\sum_{n,\textbf{k}}f_{n\textbf{k}}\Omega^n_{xy}(\textbf{k}),
\end{equation}
where $\Omega^n_{xy}(\textbf{k})$ is the band- and momentum-resolved Berry curvature, given by,
\begin{equation}\label{eq:Berry}
\Omega^n_{xy}(\textbf{k})=-\sum_{n'\neq n}\frac{2 {\rm Im}[\langle\psi_{n\textbf{k}}|\hat{v}_x|\langle\psi_{n'\textbf{k}}\rangle\langle\psi_{n'\textbf{k}}|\hat{v}_y|\langle\psi_{n\textbf{k}}\rangle]}{(\varepsilon_{n\textbf{k}}-\varepsilon_{n'\textbf{k}})^2}.
\end{equation}
Due to the band crossings, the Berry curvature can be sensitive to the $k$-mesh. After carefully examined the convergence of intrinsic AHC, a $k$-mesh of $350\times350\times1$ is ensured to obtain reliable results (see Fig.~\textcolor{blue}{S1} in Supplementary Material).

The transverse charge current can be induced by a longitudinal temperature gradient field (instead of a longitudinal electric field), which is referred as the ANE~\cite{Nernst1887}. In general, the transverse thermal current concomitantly arise under the longitudinal temperature gradient field, called ATHE or anomalous Righi-Leduc effect~\cite{QinTao2011}.  The ANE and ATHE are usually regarded as the thermoelectric counterpart and the thermal analogue of the AHE, respectively.  The AHE, ANE, and ATHE can be physically related to each other via the anomalous transport coefficients in the generalized Landauer-B\"uttiker formalism~\cite{ashcroft1976solid,Houten1992,behnia2015fundamentals},
\begin{equation}\label{eq:LB}
R^{(n)}_{xy}=\int^\infty_{-\infty}d\varepsilon(\varepsilon-\mu)^n(-\frac{\partial f}{\partial\varepsilon})\sigma^\textnormal{T=0}_{xy}(\varepsilon),
\end{equation}
where $\varepsilon$, $\mu$, and $f$ are energy, chemical potential, and Fermi-Dirac distribution function, respectively.  Then, the temperature-dependent AHC ($\sigma^\textnormal{A}_{xy}$), ANC ($\alpha^\textnormal{A}_{xy}$) and ATHC ($\kappa^\textnormal{A}_{xy}$) read
\begin{equation}\label{eq:AHC_2}
\sigma^\textnormal{A}_{xy}=R^{(0)}_{xy},
\end{equation}
\begin{equation}\label{eq:ANC_2}
\alpha^\textnormal{A}_{xy}=-R^{(1)}_{xy}/eT,
\end{equation}
\begin{equation}\label{eq:ATHC}
\kappa^\textnormal{A}_{xy}=R^{(2)}_{xy}/e^2T.
\end{equation}

The first-principles calculations were performed using the projector augmented wave method~\cite{Blochl1994}, implemented in Vienna \textit{ab initio} simulation package (\textsc{vasp})~\cite{Kresse1993,Kresse1996}.  The generalized gradient approximation with the Perdew-Burke-Ernzerhof parameterization~\cite{Perdew1996} was utilized to treat the exchange-correlation functional.  The plane-wave cutoff energy was set to be 400 eV.  For the Brillouin zone integration, we used an $18\times18\times1$ $k$-mesh.  The convergence of total energy with respect to the cutoff energy and $k$-mesh has been well examined, and the results of MAE are reliable (see Tabs.~\textcolor{blue}{S1} and~\textcolor{blue}{S2} in Supplementary Material).  The vacuum layer with the thickness of at least 25 {\AA} for all structures was used to avoid interactions between adjacent layers and the DFT-D3 method~\cite{Grimme2010,Grimme2011} was adopted to treat the vdW correction of BL systems.  The spin-orbit coupling was included in the calculations of MAE, band structures, MOE, AHE, ANE, and ATHE.  The maximally localized Wannier function
s were obtained by \textsc{WANNIER90} package~\cite{Arash2008}, where the $p$-orbitals of Ge and Te atoms as well as the $d$-orbitals of Fe atoms were projected onto the Wannier functions.  The band structures obtained from Wannier-interpolation are identical to the DFT ones around the Fermi level, as shown in supplementary Figs.~\textcolor{blue}{S2}--\textcolor{blue}{S4}.  To obtain the magnetic point and space groups of various magnetic structures, the ISOTROPY software~\cite{isotropy} is used, in which the lattice constants, atomic positions, and the direction and size of spin magnetic moment on each atom are necessary input parameters.

\begin{figure*}
	\includegraphics[width=2\columnwidth]{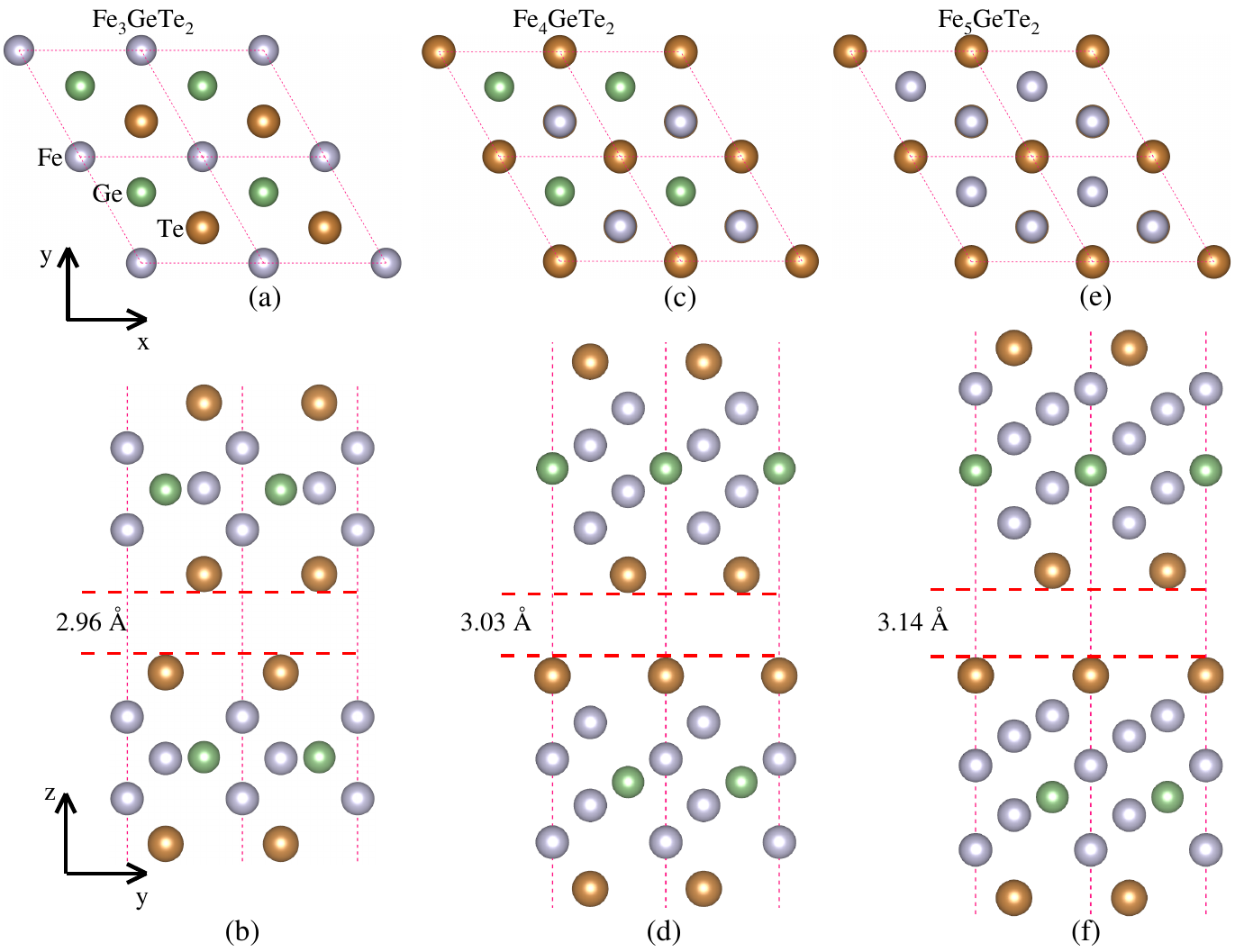}
	\caption{(Color online) (a,c,e) Top views of monolayer Fe$_n$GeTe$_2$ ($n$ = 3, 4, 5).  The silver, green, and orange spheres represent Fe, Ge, and Te atoms, respectively.  The pink dashed lines draw up the 2D primitive cell.  (b,d,f) Side views of bilayer Fe$_n$GeTe$_2$ ($n$ = 3, 4, 5).  The interlayer vdW gaps of Fe$_3$GeTe$_2$, Fe$_4$GeTe$_2$, and Fe$_5$GeTe$_2$ are 2.96 {\AA}, 3.03 {\AA}, and 3.14 {\AA}, respectively.}
	\label{fig:crystal}
\end{figure*}

\begin{figure*}
	\includegraphics[width=2\columnwidth]{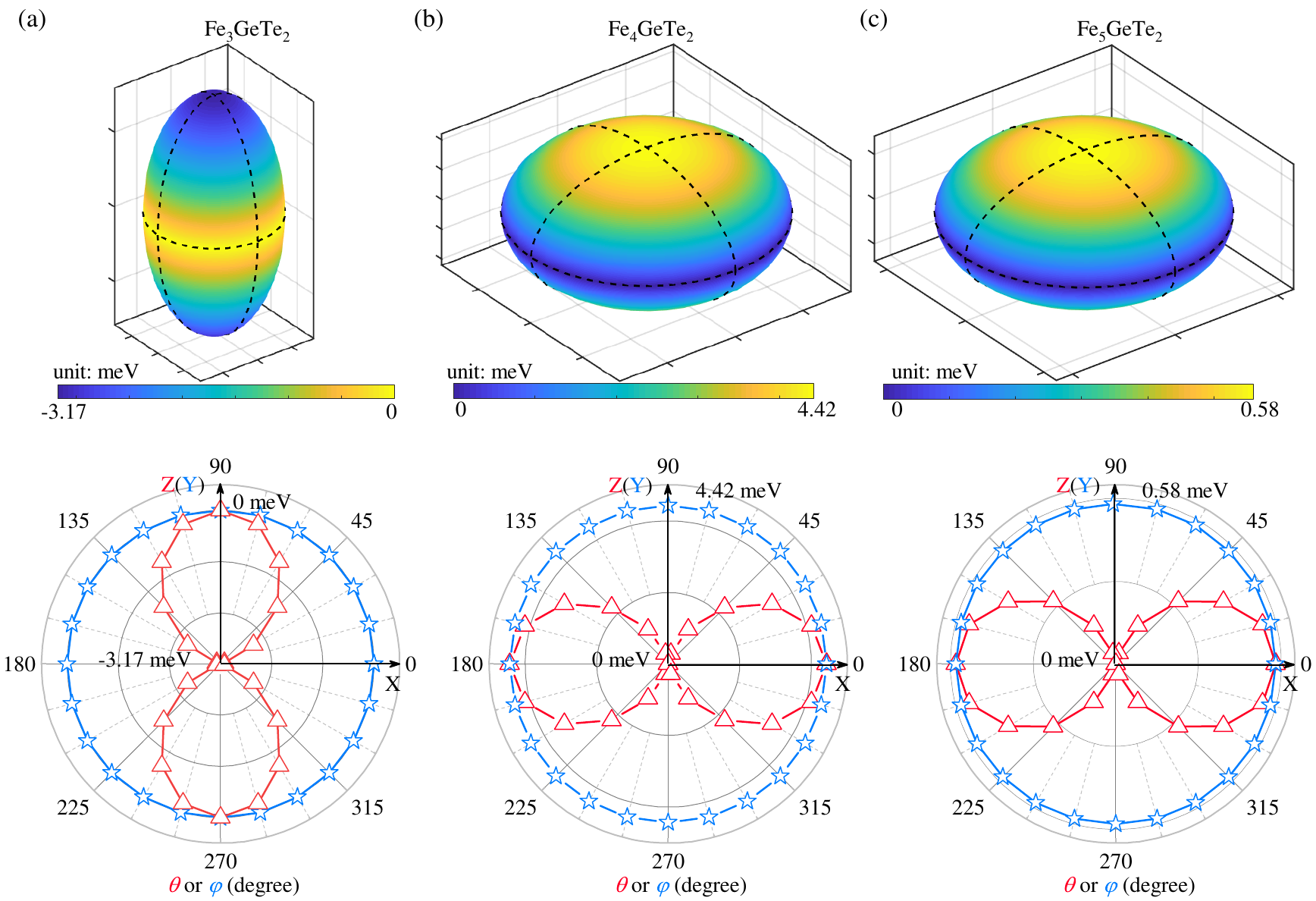}
	\caption{(Color online)  The calculated magnetocrystalline anisotropy energy of monolayer Fe$_n$GeTe$_2$ ($n$ = 3, 4, 5).  Top patterns are the three-dimensional mapping of MAE, whereas bottom patterns are the extracted data from the equator and longitude lines (black dashed lines) of the MAE map.  Apparently, F3GT favors the out-of-plane magnetization, while F4GT and F4GT present the in-plane magnetization.}
	\label{fig:MAE}
\end{figure*}

\section{Results and discussion}\label{results}
In this section, we shall present our results as follows.  First, the crystal structures of ML and BL Fe$_n$GeTe$_2$ ($n$ = 3, 4, 5) are discussed and the magnetic ground states are confirmed by the calculations of MAE.  Then, utilizing magnetic group theory, the shape (nonzero off-diagonal elements) of optical conductivity tensor is determined and hence the symmetry requirements for the MOE and anomalous transport properties (AHE, ANE, and ATHE) are obtained.  Subsequently, the symmetrically-allowed MOE, AHE, ANE, and ATHE for ML and BL ferromagnetic Fe$_n$GeTe$_2$ are calculated by the first-principles methods.  Finally, the BL antiferromagnetic Fe$_5$GeTe$_2$ that surprisingly generates the MOE, AHE, ANE, and ATHE is detailedly discussed as a special case.

\subsection{Crystal, magnetic and electronic structures}
The top and side views of the vdW layered magnets Fe$_n$GeTe$_2$ ($n$ = 3, 4, 5) with hexagonal structure exfoliating from bulk crystals~\cite{Fei2018,Deng2018,May2019,Seo2020} are shown in Fig.~\ref{fig:crystal}.  The optimized lattice constants and atomic positions of ML and BL F3GT, F4GT, and F5GT are listed in Tab.~\ref{tab:MAE_tab} and supplementary Tabs.~\textcolor{blue}{S3}--\textcolor{blue}{S8}.  Each F3GT (F4GT, F5GT) ML is consisted of three (four, five) Fe atoms, one Ge atom, and two Te atoms.  From Figs.~\ref{fig:crystal}(a) and~\ref{fig:crystal}(c), one can see that for ML F3GT and F4GT, every three identical atoms form a triangle and other atoms are located in the center of the triangle.  However, for ML F5GT, six Fe atoms form a hexagon.  The calculated effective thickness of ML F3GT, F4GT, and F5GT are 8.13 {\AA}, 9.42 {\AA}, and 9.83 {\AA}, respectively, which are consistent with the experimental data~\cite{Fei2018,Deng2018,May2019,Seo2020}.  To investigate bilayer structures, the stacking order should be primarily determined.  As shown in supplementary Fig.~\textcolor{blue}{S5} and Tab.~\textcolor{blue}{S9}, we constructed different stacking orders of BL Fe$_n$GeTe$_2$ and calculated the total energies of ferromagnetic and antiferromagnetic structures.  Our results show that the stacking orders taken from experimental bulk crystal structures, as shown in Figs.~\ref{fig:crystal}(b),~\ref{fig:crystal}(d), and~\ref{fig:crystal}(f), have the lowest energies.  The relaxed interlayer vdW gaps of BL F3GT, F4GT, and F5GT are 2.96 {\AA}, 3.03 {\AA}, and 3.14 {\AA}, respectively.  The vdW gap of F3GT is consistent with the experimental data (2.95 {\AA})~\cite{Deng2018}.  The large vdW gaps of F4GT and F5GT indicate their weak interlayer coupling, and it is likely to obtain atomically thin structures (ML or BL) by mechanically exfoliating from bulk crystals, similarly to F3GT~\cite{Deng2018,Fei2018}. 

As one knows, the magnetic anisotropy, which is a preferred direction for spin aligning, is beneficial to stabilize the long-range magnetic order in 2D systems.  The magnetic anisotropy has various sources, such as shape anisotropy, stress anisotropy, and the foremost magnetocrystalline anisotropy~\cite{Gong2017}.  The MAE is usually defined as the difference of total energy between two different spin directions, e.g., $\Delta E_{\textnormal{MAE}}(\theta,\varphi)=E(\theta,\varphi)-E(\theta=90^\circ,\varphi=0^\circ)$, here $E(\theta,\varphi)$ is the total energy when the spin points to the direction labeled by the polar ($\theta$) and azimuthal ($\varphi$) angles.  Positive $\Delta E_{\textnormal{MAE}}$ indicates an favored in-plane magnetization along the $x$-axis ($\theta=90^\circ$, $\varphi=0^\circ$), and negative $\Delta E_{\textnormal{MAE}}$ indicates the out-of-plane magnetization.  The MAE of ML and BL F3GT, F4GT, and F5GT are shown in Fig.~\ref{fig:MAE} and Tab.~\ref{tab:MAE_tab}.  Fig.~\ref{fig:MAE} illustrates that ML F3GT is an Ising ferromagnet since there exists an easy axis along the $z$-direction, while ML F4GT and F5GT belong to the category of XY ferromagnets due to the magnetic isotropy on the $xy$ plane (i.e., the easy plane)~\cite{ZhuangHL2016}.  Moreover, in the $zx$ plane the angular dependence of MAE can be fitted well by the equation, MAE($\theta$) = $K_1\cos^2 \theta + K_2\cos^4 \theta$, where $K_1$ and $K_2$ are the magnetocrystalline ansiotropy coefficients.  The fitted coefficients (in unit of meV) for ML F3GT, F4GT, and F5GT are $K_1=-2.59$ and $K_2=-0.58$, $K_1=4.48$ and $K_2=-0.06$, and $K_1=-0.67$ and $K_2=-0.09$, respectively.  The maximum values of MAE in ML F3GT, F4GT, and F5GT between the out-of-plane ($z$-axis) and in-plane ($x$-axis)  magnetization reach to -3.17, 4.42 and 0.58 meV/f.u., respectively, which are larger than that of famous 2D ferromagnets CrI$_3$ (0.5 meV/f.u.~\cite{Kumar2019}, 0.305 meV/f.u.~\cite{LuXb2020}) and Cr$_2$Ge$_2$Te$_6$ (0.4 meV/f.u.)~\cite{FangY2018}, suggesting that ML F3GT and F4GT have appreciable applications in magnetic data storage.  Additionally, our calculated MAE value of ML F3GT is close to the experimental measurement (2.0 meV/f.u)~\cite{Deng2018} and is consistent with the previously theoretical calculation (3.0 meV/f.u.)~\cite{JiangMC2020}.  In contrast to the out-of-plane magnetization of ML F3GT, ML F4GT and F5GT exhibit the in-plane magnetization, as shown in Figs.~\ref{fig:MAE}(a-c). However, in recent experimental reports the nanometer-thick F4GT and F5GT samples show the out-of-plane magnetic anisotropy~\cite{May2019,Seo2020,Ohta_2020}.  The difference between computational and experimental results can be explained by the thickness of F4GT and F5GT thin-films.  In our calculations, F3GT, F4GT, and F5GT are ML structures, whereas the e xperimental samples are multilayer structures (the thinnest sample is seven-layer for F4GT~\cite{Seo2020} and five-layer for F5GT~\cite{Ohta_2020}).  Additionally, the composition ratio of Fe, Ge, and Te atoms in experimental samples is not strictly to be 4:1:2~\cite{Seo2020} or 5:1:2~\cite{May2019,Ohta_2020}, which may also affect the magnetic ground states.  From Tab.~\ref{tab:MAE_tab}, one can observe that the MAE of F3GT decreases with increasing number of layers, which is in good agreement with other theoretical work~\cite{JiangMC2020}.  This trend also appears in F4GT and F5GT, suggesting the out-of-plane magnetic anisotropy enhances when the number of layers increases.  Therefore, a reasonable guess is that with increasing number of layers, multilayer F4GT and F5GT eventually exhibit the out-of-plane magnetization.  To verify, we further calculate the MAE of bulk F4GT and F5GT, listed in Tab.~\ref{tab:MAE_tab}, from which one can indeed find that the bulk F4GT and F5GT prefer to out-of-plane magnetization.

The crystal structures of BL Fe$_n$GeTe$_2$ are plotted in bottom row of Figs.~\ref{fig:crystal}.  We further consider two magnetic structures of BL Fe$_n$GeTe$_2$ which have ferromagnetic or antiferromagnetic interlayer coupling.  The calculated energy difference and magnetic ground states are summarized in Tab~\ref{tab:MAE_tab}, from which one can see that BL F4GT prefers to be ferromagnetism, while BL F3GT and F5GT are favorable to be antiferromagnetism.  However,  BL antiferromagnetic F3GT is extremely fragile to dopants or defects~\cite{jang2019origin}, and the ferromagnetic phase of BL F3GT was observed by experiments~\cite{Fei2018,Deng2018}.  The BL ferromagnetic F4GT has been confirmed and the intralayer ferromagnetic coupling of F4GT is stronger than F3GT due to the two Fe-Fe dumbbells~\cite{Seo2020}.  Because of tunable Fe contents and Fe vacancies, F5GT exhibits more complex magnetic behaviors than F3GT~\cite{ZhangHR2020}.  Additionally, the crystal quality of F5GT strongly depends on how the samples are cooled down~\cite{Ohta_2020}.  From Tab.~\ref{tab:MAE_tab}, one can see that the perfect BL F5GT shows the interlayer antiferromagnetic coupling.  Compared with ferromagnets, antiferromagnets are more advantageous for spintronics and spin caloritronics because of robustness against magnetic perturbation, negligible stray field, and ultrafast spin dynamics~\cite{BaltzV2018}.  Therefore, BL antiferromagnetic F5GT may be more applicable to spintronics and spin caloritronics.

\begin{table}[htpb]\footnotesize
	\caption{The calculated lattice constants ($a$), MAE ($\Delta E_{\textnormal{MAE}}$), and magnetic ground states (MGS) of monolayer and bilayer Fe$_n$GeTe$_2$ ($n$ = 3, 4, 5).  Here, $\Delta E_{\textnormal{MAE}} = E(\theta=0^\circ,\varphi=0^\circ)-E(\theta=90^\circ,\varphi=0^\circ)$ and positive (negative) value indicates a favored in-plane (out-of-plane) magnetization. The energy difference between ferromagnetic and antiferromagnetic bilayer Fe$_n$GeTe$_2$, $\Delta E_{\textnormal{FM-AFM}}$, are given in the parentheses, where positive (negative) value indicates that AFM (FM) structure is energetically stable.  The units of $\Delta E_{\textnormal{MAE}}$ and $\Delta E_\textnormal{FM-AFM}$ are meV/f.u. and the unit of $a$ is {\AA}.}
	\label{tab:MAE_tab}
	\begin{ruledtabular}
		\begingroup
		\setlength{\tabcolsep}{4.5pt} % Default value: 6pt
		\renewcommand{\arraystretch}{1.5} % Default value: 1
		\begin{tabular}{lcccl}
			
       & &$a$ &$\Delta E_{\textnormal{MAE}}$ &MGS ($\Delta E_\textnormal{FM-AFM}$)  \\
		\hline
		F3GT & ML & 4.02 & -3.17  & FM    \\
             & BL  & 4.01 & -3.23  & AFM (2.89)  \\
             & Bulk  & 4.04 & -3.25  &   \\
		\hline
        F4GT & ML & 3.92 & 4.42  & FM   \\
		     & BL  & 3.93 & 1.98  & FM (-13.48)   \\
		     & Bulk &  3.97 &-1.33 \\
		\hline
		F5GT & ML & 3.96 & 0.58  & FM   \\
		     & BL  & 3.97 & 0.32  & AFM (0.83)  \\
		     & Bulk & 4.00 &-0.55 \\
		\end{tabular}
		\endgroup
	\end{ruledtabular}
\end{table}

\begin{figure*}
	\includegraphics[width=2\columnwidth]{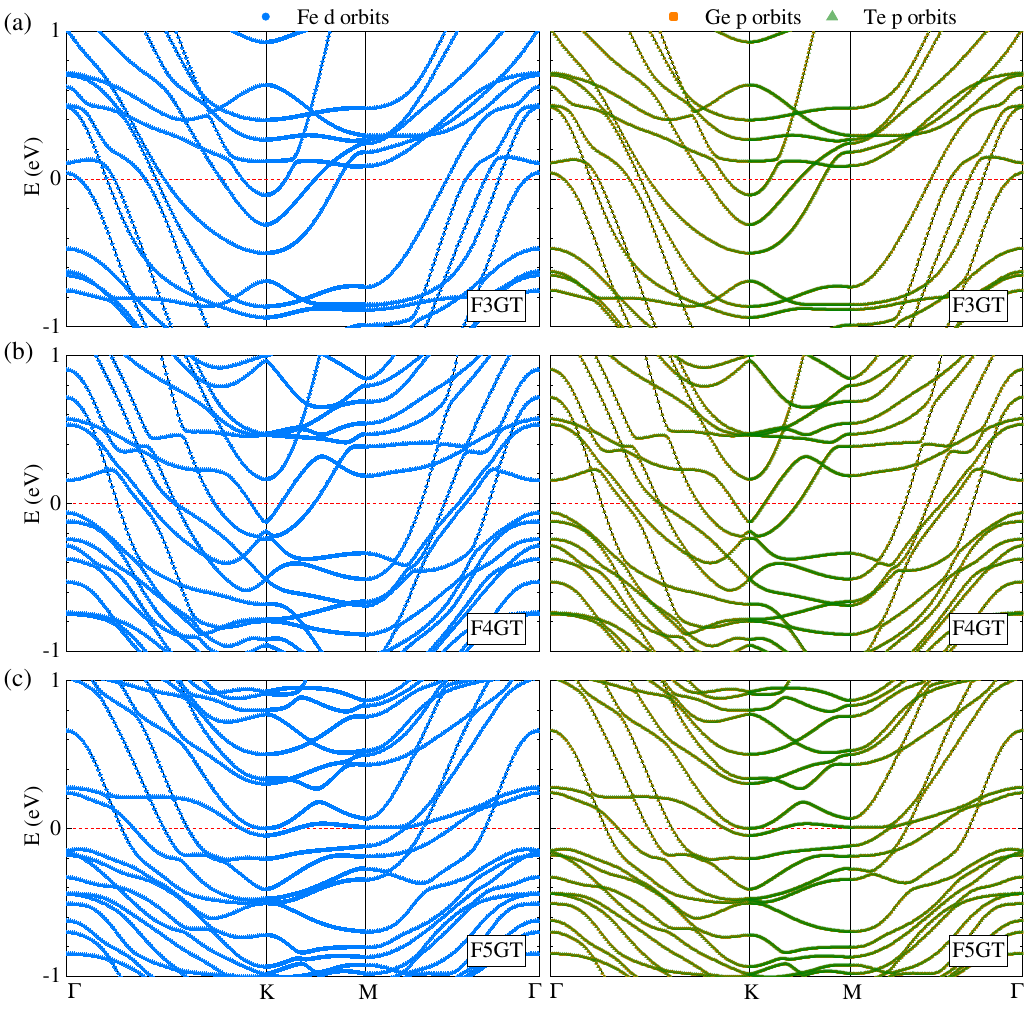}
	\caption{(Color online)  Orbital-projected band structures of monolayer ferromagnetic Fe$_n$GeTe$_2$ ($n$ = 3, 4, 5) with the out-of-plane magnetization.  The blue, orange, and green represent the $d$-, $p$-, and $p$-orbitals of Fe, Ge, and Te atoms, respectively.}
	\label{fig:band}
\end{figure*}

We then discuss the electronic structures of Fe$_n$GeTe$_2$ ($n$ = 3, 4, 5) by taking the ML ferromagnetic states with the out-of-plane magnetization as examples, as plotted in Fig.~\ref{fig:band}.   Around the Fermi level, the $d$-, $p$-, and $p$-orbitals of Fe, Ge, and Te atoms, respectively, are dominant components.  Moreover, the $p$-orbitals of Ge and Te atoms have nearly equal contributions.  One can see that ML ferromagnetic Fe$_n$GeTe$_2$ are metals with a complex multi-band character at the Fermi level.  ML ferromagnetic Fe$_n$GeTe$_2$ with the in-plane  magnetization as well as BL ferromagnetic and antiferromagnetic Fe$_n$GeTe$_2$ with both out-of-plane and in-plane magnetization are also multi-band metals, see supplementary Figs.~\textcolor{blue}{S2}--\textcolor{blue}{S4}.

\subsection{Magnetic group theory}

Before practically calculating the MOE, AHE, ANE, and ATHE, a preceding step is to find their symmetry requirements.  According to Eqs.~\eqref{eq:LB}-\eqref{eq:ATHC}, the finite-temperature AHC, ANC, and ATHC can be obtained from the zero-temperature AHC.  Additionally, the dc limit of the real part of the off-diagonal element of optical conductivity is just the zero-temperature AHC [Eqs.~\eqref{eq:OPC}-\eqref{eq:AHC}].  Therefore, the AHE, ANE, and ATHE as well as MOE should share the same symmetry requirements.  And we only need to study the shape (nonzero off-diagonal elements) of optical conductivity tensor ($\sigma$). Magnetic group theory is a powerful tool to identify nonvanishing elements of $\sigma$.   Because of the translationally invariance of $\sigma$, it is sufficient to restrict our analysis to magnetic point group.  Here, the magnetic point groups of various magnetic structures are calculated by ISOTROPY software~\cite{isotropy} and the results are summarized in Tab.~\ref{tab:group} and supplementary Tab.~\textcolor{blue}{S10}.  Additionally, the off-diagonal elements of $\sigma$ can be regarded as a pseudovector, like spin, and therefore its vector-form notation, $\sigma = [\sigma^x, \sigma^y, \sigma^z] = [\sigma_{yz}, \sigma_{zx}, \sigma_{xy}]$, is used for convenience.  It should be mentioned that in 2D systems $\sigma^x$ and $\sigma^y$ are always equal to zero, and only $\sigma^z$ ($\equiv\sigma_{xy}$) is potentially nonzero.

\begin{table*}[htbp]\footnotesize
    \caption{Magnetic point groups of monolayer and bilayer Fe$_n$GeTe$_2$ ($n$ = 3, 4, 5) when the spin points to the $x$-, $y$-, and $z$-axes.  We consider monolayer structures with only ferromagnetic order, but bilayer structures with both ferromagnetic and antiferromagnetic orders.  Y (yes) and N (no) indicate that there exists nonzero $\sigma^z$ or not.}
    \label{tab:group}
	\begin{ruledtabular}
		\begingroup
		\setlength{\tabcolsep}{4.5pt} % Default value: 6pt
		\renewcommand{\arraystretch}{1.5} % Default value: 1
    
    \begin{tabular}{lrrrrrrrrr}
                &\multicolumn{3}{c}{ML (FM)}        &\multicolumn{3}{c}{BL (FM)}        &\multicolumn{3}{c}{BL (AFM)}  \\
          \hline
          &\multicolumn{1}{c}{$x$}     &\multicolumn{1}{c}{$y$}     &\multicolumn{1}{c}{$z$}     &\multicolumn{1}{c}{$x$}     &\multicolumn{1}{c}{$y$}     &\multicolumn{1}{c}{$z$}     &\multicolumn{1}{c}{$x$}     &\multicolumn{1}{c}{$y$}     &\multicolumn{1}{c}{$z$} \\
          \hline
    F3GT  & $m'm2'$ (N) & $m'm'2$ (N) &$\bar{6}m'2'$ (Y)      & $2/m$ (N)  & $2'/m'$ (Y) & $\bar{3}1m'$ (Y)    & $2'/m$ (N)  & $2/m'$ (N)    & $\bar{3}'1m'$ (N)\\
    F4GT  & $2/m$ (N)  & $2'/m'$ (Y) &$\bar{3}1m'$ (Y)       & $2/m$ (N) & $2'/m'$ (Y) & $\bar{3}1m'$ (Y)    & $2'/m$ (N) & $2/m'$ (N)     & $\bar{3}'1m'$ (N)\\
    F5GT  & $m$ (N)    & $m'$ (Y)    & $3m'1$ (Y)      & $m$ (N)   & $m'$ (Y)    & $3m'1$ (Y)     & $m$ (N)    & $m'$ (Y)     & $3m'1$ (Y) \\
		\end{tabular}
		\endgroup
	\end{ruledtabular}
\end{table*}%

The magnetic point groups of ML Fe$_n$GeTe$_2$ ($n$ = 3, 4, 5) as a function of polar ($\theta$) and azimuthal ($\varphi$) angles when the spin rotates within the $xz$ and $xy$ planes are summarized in supplementary Tab.~\textcolor{blue}{S10}.  In the main text, we only discuss three special directions, namely, when the spin points to the $x$-axis ($\theta = 90^\circ, \varphi = 0^\circ$), $y$-axis ($\theta = 90^\circ, \varphi = 90^\circ$), and $z$-axis ($\theta = 0^\circ, \varphi = 0^\circ$), as listed in Tab~\ref{tab:group}.   Let us start with analyzing the magnetic point groups of ML F3GT.  The magnetic point groups of ML F3GT are $m'm2'$, $m'm'2$, and $\bar{6}m'2'$ when the spin points to $x$-, $y$-, and $z$-axes, respectively.  First, group $m'm2'$ has one mirror plane $\mathcal{M}_x$, which is perpendicular to the spin orientation ($x$-axis) and is parallel to the $z$-axis.  Such a mirror operation reverses the sign of $\sigma^z$, thus suggesting $\sigma^z = 0$.  Group $m'm'2$ contains one combined symmetry $\mathcal{TM}_z$, where $\mathcal{T}$ is the time-reversal symmetry and $\mathcal{M}_z$ is parallel to the spin orientation ($y$-axis) and is perpendicular to $z$-axis.  Note that $\mathcal{TM}_z$ operation reverses the sign of $\sigma^z$, and hence $\sigma^z = 0$.  When the spin is along the $z$-axis, ML F3GT belongs to magnetic point group of $\bar{6}m'2'$.  This group has a mirror plane $\mathcal{M}_z$, which is perpendicular to the spin orientation ($z$-axis) and does not change the sign of $\sigma^z$.  On the other hand, this group contains three combined symmetries ($\mathcal{TM}$), one of which is $\mathcal{TM}_x$.  The other two mirror planes can be obtained by the $\mathcal{C}_3$ symmetry.  Both the $\mathcal{T}$ and $\mathcal{M}$ operations reverse the sign of $\sigma^z$, indicating that $\sigma^z$ is even under the $\mathcal{TM}$ symmetry.  Overall, $\sigma^z$ can exist in the magnetic point group of $\bar{6}m'2'$, that is, $\sigma=[0, 0, \sigma^z]$.

The magnetic point groups of ML F4GT are $2/m$, $2'/m'$, and $\bar{3}1m'$ when the spin points to the $x$-, $y$-, and $z$-axes, respectively.  The group $2/m$ has a mirror plane $\mathcal{M}_x$, indicating $\sigma^z = 0$.  For group $2'/m'$, there is a combined symmetry $\mathcal{TM}_x$.  Both the $\mathcal{T}$ and $\mathcal{M}_x$ operations reverse the sign of $\sigma^z$, indicating $\sigma^z$ is even under $\mathcal{TM}_x$ symmetry, and hence $\sigma^z\neq0$.  The group $\bar{3}1m'$ contains three combined symmetries $\mathcal{TM}$, which are the same as group $\bar{6}m'2'$.  The combined symmetries do not change the sign of $\sigma^z$, giving rise to the nonzero $\sigma^z$.  Next, we analyze the magnetic point groups of ML F5GT, which are $m$, $m'$, and $3m'1$ when the spin is along the $x$-, $y$-, and $z$-axes, respectively.  There is a mirror plane $\mathcal{M}_x$ in group $m$, reversing the sign of $\sigma^z$, and thus $\sigma^z$ should be zero.  Group $m'$ includes a combined symmetry $\mathcal{TM}_x$, which keeps $\sigma^z\neq0$.  Group $3m'1$ has three combined symmetries $\mathcal{TM}$, like groups $\bar{6}m'2'$ and $\bar{3}1m'$, giving rise to the nonzero $\sigma^z$.

Finally, we discuss the magnetic point groups of BL ferromagnetic and antiferromagnetic Fe$_n$GeTe$_2$ ($n$ = 3, 4, 5).  It should be noted that BL F3GT and F4GT share the same magnetic point groups, as listed in Tab.~\ref{tab:group}.  The magnetic point groups of BL ferromagnetic F3GT and F4GT are $2/m$, $2'/m'$, and $\bar{3}1m'$ when the spin points to the $x$-, $y$-, and $z$-axes, respectively.  As previously mentioned, we conclude that the nonzero $\sigma^z$ can exist when the spin points to the $y$- and $z$-axes, while $\sigma^z=0$ if the spin is along $x$-axis.  The magnetic point groups of BL antiferromagnetic F3GT and F4GT are $2'/m$, $2/m'$ and $\bar{3}'1m'$, all of which have a combined symmetry $\mathcal{TP}$ ($\mathcal{P}$ is the spatial inversion symmetry), which does not allow $\sigma^z$.  Additionally, ML and BL F5GT share the same magnetic point groups when the spin is along the $x$-, $y$-, and $z$-axes.  Therefore, from our group theory analysis, $\sigma^z$ exists well in antiferromagnetic BL, just like ferromagnetic ML and BL.  Recently, there has been a consensus that $\sigma^z$ can exist in some noncollinear antiferromagnets even though the net magnetization is zero, presenting attractive MOE and anomalous transport phenomena~\cite{FengWX2015,Nayak2016,Guo_anti2017,Ikhlas2017,Lixk2017,Jungwirth2018,Libor2018,Zelezny2018,Kaori2019,XuLC2020,Zhou2020,Zhangyang2017,zhou2019Mn,dWumingxing2020,feng2020,Libor2021}.  However, it is relatively rare in collinear antiferromagnets~\cite{Zhouxiaodong2021,Smejkal2020}.  Therefore, BL antiferromagnetic F5GT may be of great interest to experimental attention.

In general, $\sigma^z$ is nonzero in ML ferromagnetic F3GT when the spin is along the $z$-axis and is also nonzero in ML ferromagnetic F4GT and F5GT, BL ferromagnetic F3GT, F4GT, and F5GT, as well as BL antiferromagnetic F5GT when the spin points to the $y$- and $z$-axes, which are all summarized in Tab.~\ref{tab:group}.  In the following, we restrict our discussion on the MOE, AHE, ANE, and ATHE of the above-mentioned magnetic structures.

\begin{figure*}
	\includegraphics[width=2\columnwidth]{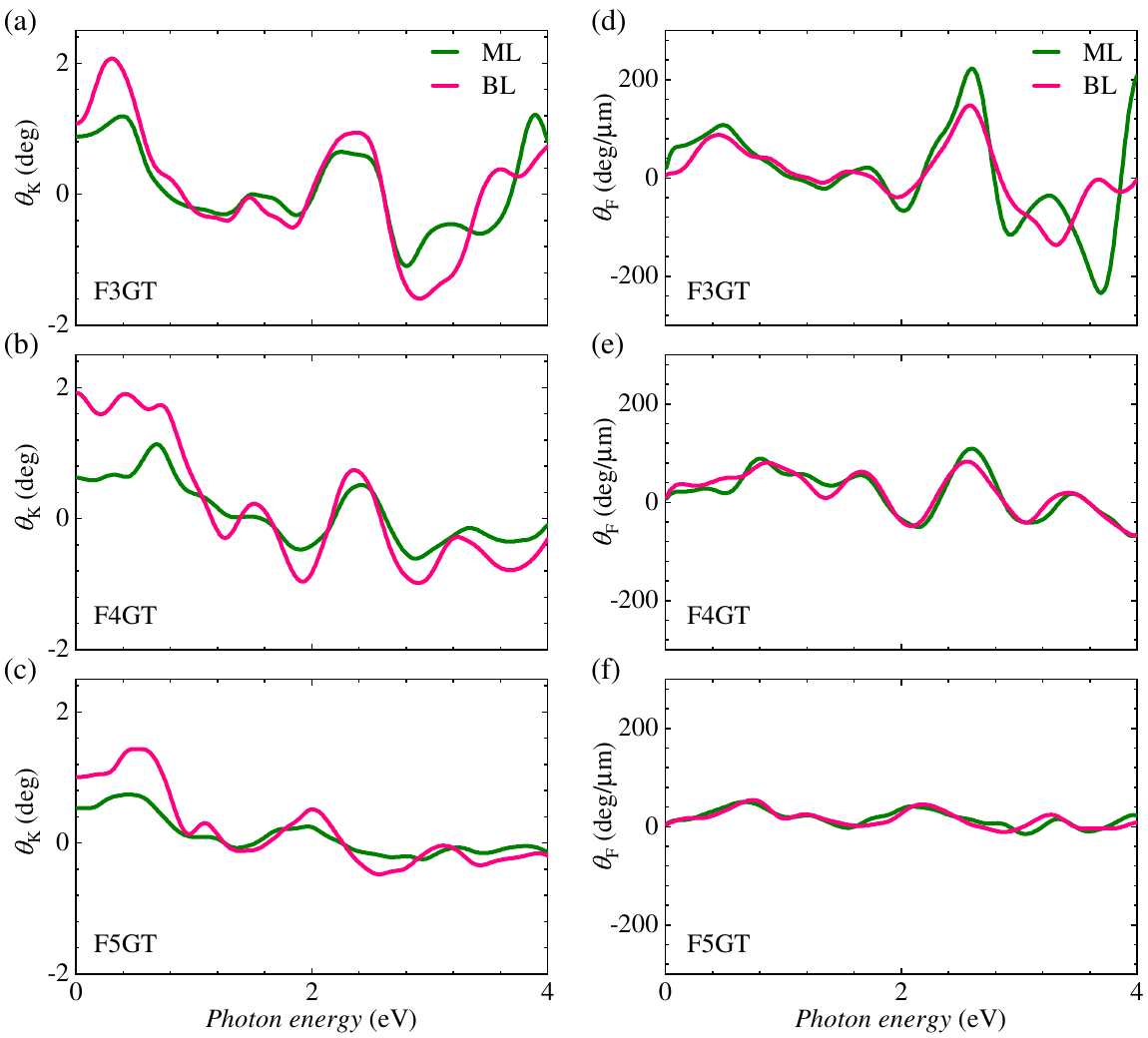}
	\caption{(Color online) The calculated magneto-optical Kerr (a-c) and Faraday (d-f) rotation angles for monolayer (olive lines) and bilayer (pink lines) ferromagnetic Fe$_n$GeTe$_2$ ($n$ = 3, 4, 5) with the out-of-plane magnetization (along the $z$-axis).}
	\label{fig:Kerr_Faraday}
\end{figure*}

\subsection{Magneto-optical effects}

The band structures of ML and BL Fe$_n$GeTe$_2$ ($n$ = 3, 4, 5) are plotted in supplementary Figs.~\textcolor{blue}{S2}--\textcolor{blue}{S4}, from which the multi-band metals can be easily confirmed.  According to Eqs.~\eqref{eq:Kerr}-\eqref{eq:Faraday2}, the magneto-optical Kerr and Faraday angles are closely related to the optical conductivity, which are shown in supplementary Figs.~\textcolor{blue}{S6}-\textcolor{blue}{S9}.  The diagonal elements of optical conductivity are nearly isotropy regardless of the spin's direction, while the off-diagonal elements of optical conductivity exhibit strong anisotropy between the in-plane and out-of-plane magnetization.  Consequently, the magneto-optical Kerr and Faraday angles also show strong anisotropy, as shown in supplementary Figs.~\textcolor{blue}{S10} and~\textcolor{blue}{S11}.  Although the MAE results show that ML and BL F4GT and F5GT prefer to the in-plane magnetization, the magnetized direction of 2D materials can be feasibly modulated by applying an external magnetic field~\cite{Zhuang2016}.  In the main text, we take the out-of-plane magnetization as an example to discuss the MOE as well as the following AHE, ANE, and ATHE.

Figure~\ref{fig:Kerr_Faraday} displays the magneto-optical Kerr and Faraday angles of ML and BL ferromagnetic Fe$_n$GeTe$_2$ with the out-of-plane magnetization (i.e., $z$-axis) as a function of photon energy.  In the energy range of $0 \sim 4$ eV, the Kerr ($\theta_\textnormal{K}$) and Faraday ($\theta_\textnormal{F}$) angles oscillate frequently.  For ML F3GT, F4GT, and F5GT, the prominent peaks of Kerr angle locate at \{0.39, 2.25, 2.80\} eV, \{0.69, 1.90, 2.42, 2.88\} eV, and \{0.45, 1.96\} eV, respectively.  For BL structures, the locations of prominent peaks appear close to that of ML structures.  Remarkably, the amplitudes of the $\theta_\textnormal{K}$ of BL structures are larger than that of ML structures.  The maximum values of $\theta_\textnormal{K}$ for ML (BL) F3GT, F4GT, and F5GT are 1.19$^\circ$ (2.07$^\circ$), 1.14$^\circ$ (1.90$^\circ$), 0.74$^\circ$ (1.44$^\circ$), respectively.  Our calculated Kerr angles of ML and BL F3GT are in agreement with the previously theoretical report (1.65$^\circ$ and 2.76$^\circ$ for ML and BL, respectively)~\cite{JiangMC2020}.  The calculated largest values of $\theta_\textnormal{F}$ for ML (BL) F3GT, F4GT, and F5GT are 222.66$^\circ/\mu m$ (147.73$^\circ/\mu m$), 109.12$^\circ/\mu m$ (82.95$^\circ/\mu m$), 50.28$^\circ/\mu m$ (54.07$^\circ/\mu m$), respectively.

The $\theta_\textnormal{K}$ of ML and BL ferromagnetic Fe$_n$GeTe$_2$ are larger than that of the most traditional ferromagnets.  For example, the $\theta_\textnormal{K}$ of bcc Fe, hcp Co, and fcc Ni are only -0.5$^\circ$, -0.42$^\circ$, and -0.25$^\circ$~\cite{Antonov2004book,Guo1995_MO,Oppeneer1992}, respectively.  Moreover, the $\theta_\textnormal{K}$ of ML and BL ferromagnetic Fe$_n$GeTe$_2$ are larger than or comparable to that of many famous 2D magnetic materials, such as CrI$_3$ [0.29$^\circ$ (ML), 2.86$^\circ$ (trilayer)]~\cite{Huang2017}, Cr$_2$Ge$_2$Te$_6$ [0.9$^\circ$ (ML), 1.5$^\circ$ (BL), 2.2$^\circ$ (trilayer)]~\cite{FangY2018}, CrTe$_2$ [0.24$^\circ$ (ML), -1.76$^\circ$ (trilayer)]~\cite{YangXX2021}, blue phosphorene [0.12$^\circ$ (ML), 0.03$^\circ$ (BL)]~\cite{Zhou2017}, gray arsenene [0.81$^\circ$ (ML), 0.14$^\circ$ (BL)]~\cite{Zhou2017}, and InS [0.34$^\circ$ (ML)]~\cite{Feng2017}.  Additionally, the $\theta_\textnormal{F}$ of ML and BL ferromagnetic Fe$_n$GeTe$_2$ are larger than or comparable to that of MnBi thin-films [80$^\circ/\mu m$]~\cite{DiGQ1996,Ravindran1999}, Cr$_2$Ge$_2$Te$_6$ [$\sim$ 120$^\circ/\mu m$ (ML)]~\cite{FangY2018}, CrI$_3$ [108$^\circ/\mu m$ (TL)]~\cite{Kumar2019}, and CrTe$_2$ [-173$^\circ/\mu m$ (ML)]~\cite{YangXX2021}.  The large magneto-optical Kerr and Faraday effects suggest that ML and BL ferromagnetic Fe$_n$GeTe$_2$ may have potential applications in various magneto-optical devices.

\begin{figure*}[htbp]
	\includegraphics[width=2\columnwidth]{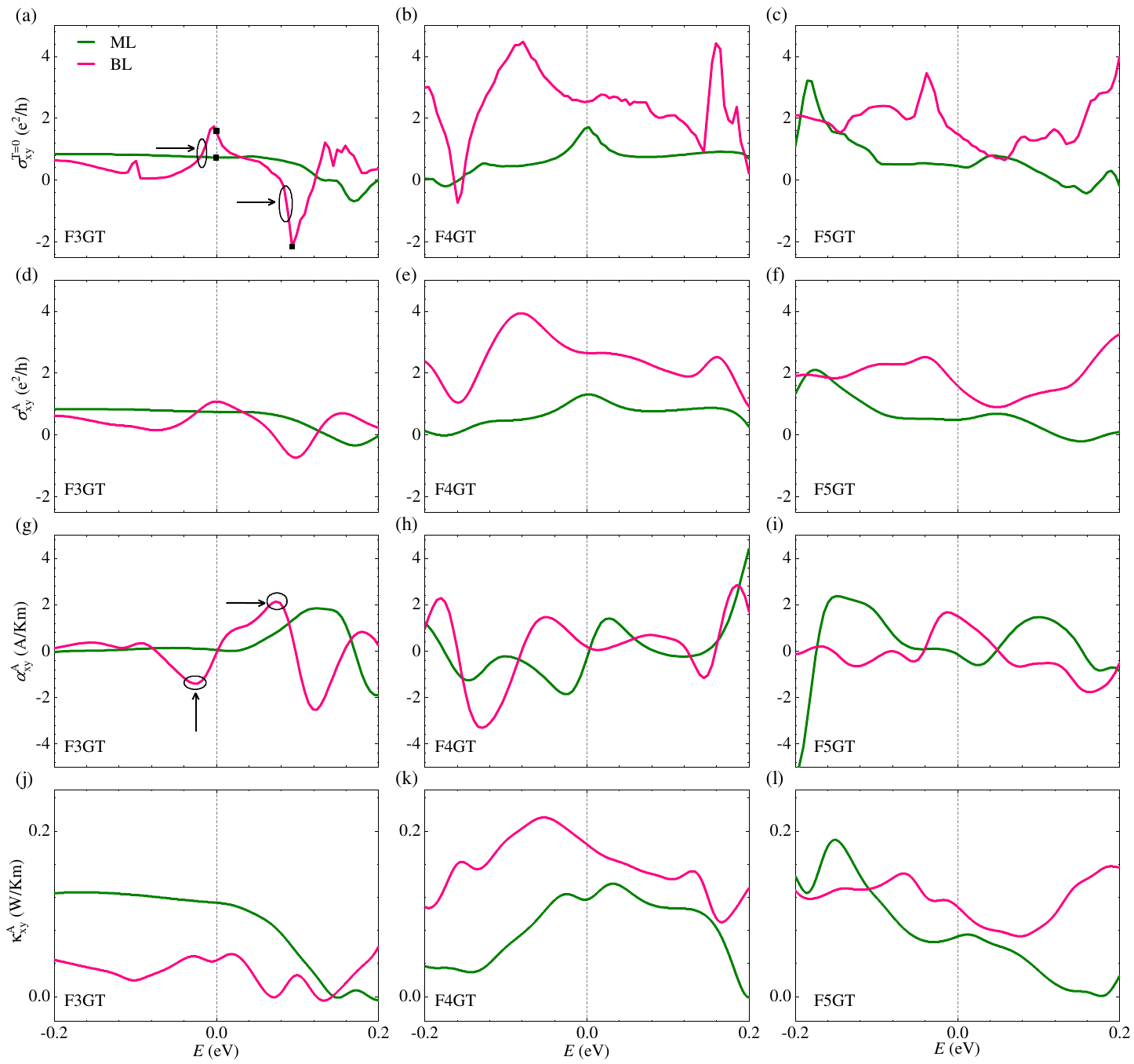}
	\caption{(Color online) The calculated anomalous Hall conductivities at 0 K (a,b,c) and 130 K (d,e,f), anomalous Nernst conductivity at 130 K (g,h,i), and anomalous thermal Hall conductivity at 130 K (j,k,l) for monolayer and bilayer ferromagnetic Fe$_n$GeTe$_2$ ($n$ = 3, 4, 5) with the out-of-plane magnetization (along the $z$-axis) as a function of the Fermi energy.  The black squares in (a) indicate the positions where the Berry curvatures shown in Fig.~\ref{fig:Berry} are calculated.  The black circles in (a) and (g) indicate the large slopes of anomalous Hall conductivity and the sharp peaks of anomalous Nernst conductivity, respectively.}
	\label{fig:AHC_ANC}
\end{figure*}

\begin{figure*}
	\includegraphics[width=2\columnwidth]{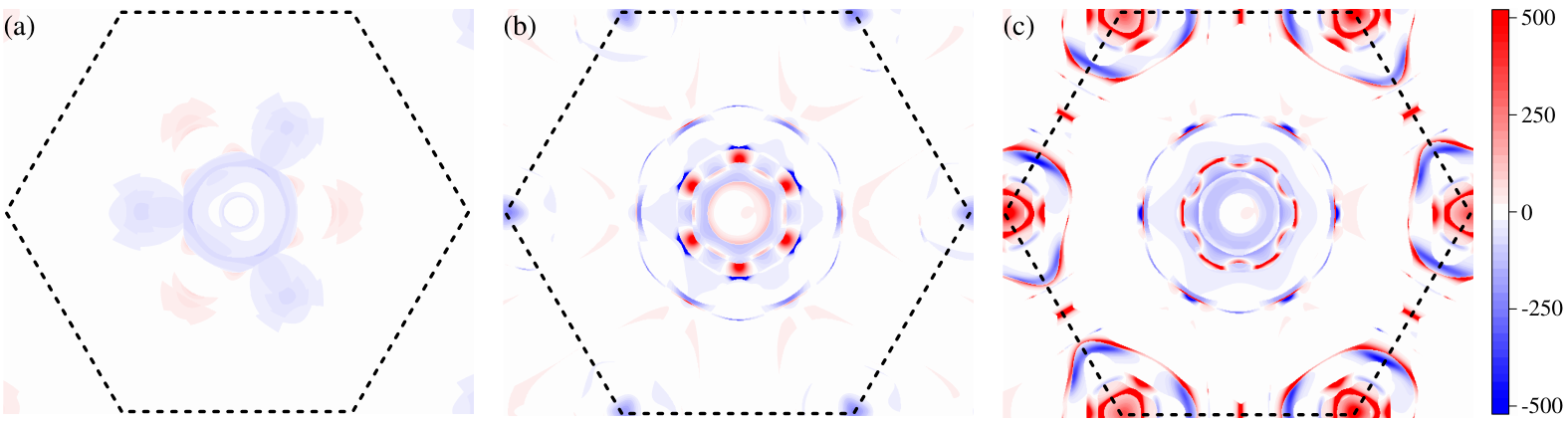}
	\caption{(Color online) The calculated momentum-resolved Berry curvature (in arbitrary units) of monolayer (a) and bilayer (b,c) Fe$_3$GeTe$_2$ on the $k_x-k_y$  plane.  The Berry curvatures in (a) and (b) are calculated at the true Fermi energy ($E_F$), while in (c) is calculated at $E_F+ 0.09$ eV.  The black dashed lines draw up the first Brillouin zone.}
	\label{fig:Berry}
\end{figure*}

\subsection{Anomalous Hall, anomalous Nernst, and anomalous thermal Hall effects}
Now, let us focus on the anomalous transport properties of ML and BL ferromagnetic Fe$_n$GeTe$_2$ ($n$ = 3, 4, 5).  The zero-temperature AHC calculated via Eq.~\eqref{eq:AHC} are plotted in Figs.~\ref{fig:AHC_ANC}(a-c) and supplementary Figs.~\textcolor{blue}{S12}(a-c) and ~\textcolor{blue}{S13}(a-c).  Similarly to the MOE, the zero-temperature AHC also exhibits large magnetic anisotropy, i.e., the results of out-of-plane magnetization are much larger than that of in-plane magnetization.  Hence, we only discuss the case of out-of-plane magnetization in the main text.  The calculated zero-temperature AHC for ML F3GT, F4GT, and F5GT are 0.73 $e^2/h$, 1.70 $e^2/h$, and 0.46 $e^2/h$ at the Fermi energy ($E_F$), respectively.  With the increasing number of layers, AHC can be greatly enhanced.  For example, the zero-temperature AHC for BL F3GT, F4GT, and F5GT increase up to 1.57 $e^2/h$, 2.54 $e^2/h$, and 1.50 $e^2/h$ at the $E_F$, respectively.  Our calculated zero-temperature AHC of ML and BL F3GT agree well with the previously theoretical report [0.37 $e^2/h$ (ML), 1.50 $e^2/h$ (BL)]~\cite{Lin2019}.  By appropriate doping electrons or holes, not only the Curie temperature can be increased, but also the magnetotransport properties can be modulated.  Intuitively speaking, doping electrons or holes can shift the position of the Fermi energy.  From Figs.~\ref{fig:AHC_ANC}(a-c), one can find that by doping electrons or holes, the AHC of ML F3GT and F4GT do not improve too much, while the AHC of ML F5GT as well as of BL F3GT, F4GT, and F5GT enhance significantly.  For example, the AHC of BL F3GT, F4GT, and F5GT increase up to -2.16 $e^2/h$, 4.48 $e^2/h$, and 3.46 $e^2/h$ when the Fermi energy shifts to 0.09 eV, -0.07 eV, and -0.04 eV, respectively.  Among Fe$_n$GeTe$_2$ ($n$ = 3, 4, 5), BL F4GT shows the largest AHC of 4.48 $e^2/h$ ($\approx$ 920 S/cm), which is larger than bulk F3GT (140--540 S/cm)~\cite{KimKyoo2018}, thin-films F3GT (360--400 S/cm)~\cite{Xu2019}, and thin-films F4GT ($\sim$ 180 S/cm)~\cite{Seo2020}, and is even compared with bcc Fe (751 S/cm~\cite{Yao2004}, 1032 S/cm~\cite{Dheer1967}).

Taking F3GT as an example, we interpret the variation of the AHC for ML and BL structures as well as for the doping effect in BL structures.  Figure~\ref{fig:Berry} plots the momentum-resolved Berry curvature of ML and BL F3GT.  At the true Fermi energy ($E_F$), the negative spots of Berry curvature are obviously larger than the positive ones [Fig.~\ref{fig:Berry}(a)], resulting in the positive value of AHC [refer to Eqs.~\eqref{eq:AHC}and~\eqref{eq:Berry} and see Fig.~\ref{fig:AHC_ANC}(a)].  Additionally, the darker red and blue colors indicate that the difference between positive and negative spots of Berry curvature becomes larger [Fig.~\ref{fig:Berry}(b)], giving rise to the larger AHC in BL F3GT than that in ML F3GT [see Fig.~\ref{fig:AHC_ANC}(a)].  In Fig.~\ref{fig:Berry}(c), when the Fermi energy is shifted to $E_F+0.09$ eV for BL F3GT, the positive spots of Berry curvature turn to dominate in the first Brillouin zone, and thus lead to a large peak of negative AHC [see Fig.~\ref{fig:AHC_ANC}(a)].

According to Eqs.~\eqref{eq:LB}-\eqref{eq:ATHC}, the temperature-dependent AHC, ANC, and ATHC can be computed from zero-temperature AHC.  The metallicity and high $T_C$ of 2D Fe$_n$GeTe$_2$ ($n$ = 3, 4, 5) have been experimentally confirmed~\cite{Deng2018,Fei2018,Seo2020,May2019}, such as the $T_C$ of $\sim$ 200 K in four-layer F3GT, of 270 K in thin-films F4GT, and of 280 K in thin-fims F5GT.  Additionally, the ML F3GT has been prepared by mechanically exfoliating from bulk materials and the $T_C$ is reported to be 130 K~\cite{Fei2018}.  In our calculations, the temperature-dependent AHC, ANC, and ATHC are uniformly calculated at 130 K, as plotted in the second, third, and fourth rows of Fig.~\ref{fig:AHC_ANC}, respectively.  A trend is that the AHC decreases with increasing of the temperature, especially for BL Fe$_n$GeTe$_2$ [compare the first and second rows of Fig.~\ref{fig:AHC_ANC}].  At the temperature of 130 K, the largest AHC of BL F3GT, F4GT, and F5GT reach to 1.08 $e^2/h$, 3.94 $e^2/h$, and 2.51 $e^2/h$ at the Fermi energies of 0.0 eV, -0.08 eV, and -0.03 eV, respectively.  Although the AHC at 130 K is smaller than the zero-temperature AHC, the evolution of AHC as a function of the Fermi energy is similar to the case of zero-temperature, and for instance, the AHC peaks roughly at the same energy.
 
The ANE is usually regarded as the thermoelectric counterpart of the AHE, which is a celebrated effect from the realm of spin caloritronics~\cite{Bauer2012,Boona2014}.  Equation~\eqref{eq:ANC_2} can be derived to the Mott relation at low temperature, which relates the ANC to the energy derivative of AHC~\cite{Xiao2006},
\begin{eqnarray}\label{eq:MottANC}
\alpha^\textnormal{A}_{xy}&=&-\frac{\pi^2}{3}\frac{k_B^2T}{e}\sigma_{xy}^A(\mu)'.
\end{eqnarray}
For example, there appears two large peaks of ANC near the $E_F$ [see black arrows in Fig.~\ref{fig:AHC_ANC}(g)], where the slopes of AHC are rather step [see black arrows in Fig.~\ref{fig:AHC_ANC}(a)].  The ANC of ML and BL F3GT are almost zero at the $E_F$. However, the ANC of ML F3GT reaches up to 1.85 A/Km at 0.12 eV by electron doping.  Moreover, the ANC of BL F3GT reaches up to 2.14 A/Km and -2.52 A/Km at 0.07 eV and 0.13 eV by electron doping, and reaches up to -1.40 A/Km at -0.03 eV by hole doping, respectively. The calculated ANC of BL F3GT is higher than thin-films F3GT ($\sim$ 0.3 A/Km)~\cite{Xu2019}.  The ANC of ML and BL F4GT are also almost zero at the $E_F$, similarly to F3GT.  After doping appropriate electrons or holes, the pronounced peaks of ANC arise.  For example, the ANC of ML F4GT can be -1.85 A/Km at -0.02 eV, and the ANC of BL F4GT appears to be -3.31 A/mK at -0.13 eV, see Fig.~\ref{fig:AHC_ANC}(h).   Figure~\ref{fig:AHC_ANC}(i) plots the ANC curves as a function of the Fermi energy for ML and BL F5GT.  The ANC of ML F5GT is almost zero at the $E_F$ but can exceed to $\pm2.0$ A/Km in the energy range of -0.15 $\sim$ -0.2 eV.  In contrast to ML F5GT, BL F5GT has a large ANC of 1.68 A/Km at the $E_F$.  Thus, to improve the thermoelectric performance of Fe$_n$GeTe$_2$, doping appropriate electrons or holes is an effective way.  The calculated ANC of ML and BL Fe$_n$GeTe$_2$ are larger than that of many popular magnetic materials, e.g., Heusler compounds (1 $\sim$ 8 A/Km)~\cite{Noky2020}, Co$_3$Sn$_2$S$_2$ ($\sim$ 2 A/Km)~\cite{Guin2019}, Ti$_2$MnAl ($\sim$ 1.31 A/Km)~\cite{Noky2018}, suggesting that ML and BL Fe$_n$GeTe$_2$ may have potential applications in thermoelectric devices.

Furthermore, we present the ATHE (or called anomalous Righi-Leduc effect)~\cite{QinTao2011}, that is, a transverse thermal current induced by a longitudinal temperature gradient field, which is normally thought to be the thermal analogue of the AHE.  The ATHC of ML and BL ferromagnetic Fe$_n$GeTe$_2$ ($n$ = 3, 4, 5) are plotted in last row of Fig.~\ref{fig:AHC_ANC}.  At the $E_F$, the calculated ATHC of ML (BL) F3GT, F4GT, and F5GT are 0.11 (0.04) W/Km, 0.12 (0.18) W/Km, and 0.07 (0.11) W/Km, respectively.  These values of ATHC can be further enhanced by doping electrons or holes.  For example, the largest ATHC in ML (BL) F3GT, F4GT, and F5GT can reach up to 0.16 (0.05) W/Km, 0.14 (0.22) W/Km, and 0.19 (0.16) W/Km, respectively.  The calculated ATHC of ML and BL ferromagnetic Fe$_n$GeTe$_2$ are larger than that of Fe$_3$Sn$_2$ ($\sim$ 0.09 W/Km at 300 K)~\cite{Zhang2021} and is slightly smaller than that of Co$_2$MnGa ($\sim$ 0.6 W/Km at room temperature)~\cite{XuLiangcai2020_2}, suggesting that 2D Fe$_n$GeTe$_2$ are excellent material platform to study the anomalous thermal transports.

\subsection{Antiferromagnetism of bilayer Fe$_5$GeTe$_2$}

The understanding of MOE and AHE is gradually developing.  In early stage, the MOE and AHE are usually assumed to be linearly proportional to net magnetization, and hence most of host materials are ferromagnetic or ferrimagnetic that have finite net magnetization.  In contrast, the antiferromagnets with zero net magnetization are expected to have neither MOE nor AHE.  Until 2014, Chen et al.~\cite{ChenHua2014} predicted that Mn$_3$Ir with noncollinear antiferromagnetic order has large AHE.  Subsequent studies have identified that the MOE and AHE can occur in similar noncollinear antiferromagnets~\cite{FengWX2015,Nayak2016,Guo_anti2017,Ikhlas2017,Lixk2017,Jungwirth2018,Libor2018,Zelezny2018,Kaori2019,XuLC2020,Zhou2020,Zhangyang2017,zhou2019Mn,dWumingxing2020,feng2020,Libor2021}.  Despite the comprehensive understanding of MOE and AHE in the above noncollinear magnetic systems, it has not been understood well in more popular collinear antiferromagnets.  A common wisdom is that the MOE and AHE are not easily activated in collinear antiferromagnets.  The reason is simple that (1) although the time-reversal symmetry $\mathcal{T}$ is broken, however, the symmetries combined $\mathcal{T}$ and some spatial symmetries $\mathcal{S}$ (translation or inversion) exist more generally in collinear antiferromagnets rather than noncollinear antiferromagnets; (2) Such combined symmetries $\mathcal{TS}$ force the MOE and AHE to be zero.  Sivadas et al.~\cite{Sivadas2016} theoretically predicted that the magneto-optical Kerr effect can occur in collinear antiferromagnetic MnPSe$_3$~\cite{Sivadas2016}, but an external electric field is needed to break all combined symmetries.  On the experimental aspect, the magneto-optical Kerr effect was observed in bilayer collinear antiferromagnetic CrI$_3$ using an external electric field~\cite{Jiang2018,HuangB2018}.  Very recently, the AHE and MOE are predicted in two collinear antiferromagnets RuO$_2$ and CoNb$_3$S$_6$ with the special crystal chirality~\cite{Smejkal2020,Zhouxiaodong2021}, which are naturally absent of any combined symmetries $\mathcal{TS}$.

In this subsection, we shall show that the MOE, AHE, ANE, and ATHE surprisingly exist in BL F5GT with collinear antiferromagnetic order, which is needless of any external conditions, such as an electric field.  This is much rare in realistic collinear antiferromagnets, especially for 2D materials.  The MAE and magnetic group theory of BL F5G have been discussed in the preceding subsections.  The nonzero Berry curvature are expected in BL antiferromagnetic F5GT (both out-of-plane and in-plane magnetization) because of the symmetry requirements, and the anomalous transport properties that originate from nonzero Berry curvature are also symmetrically allowed.  The band structure plotted in supplementary Figs.~\textcolor{blue}{S4}(e) and~\textcolor{blue}{S4}(f) clearly show the metallic character of BL antiferromagnetic F5GT.

\begin{figure}
	\includegraphics[width=\columnwidth]{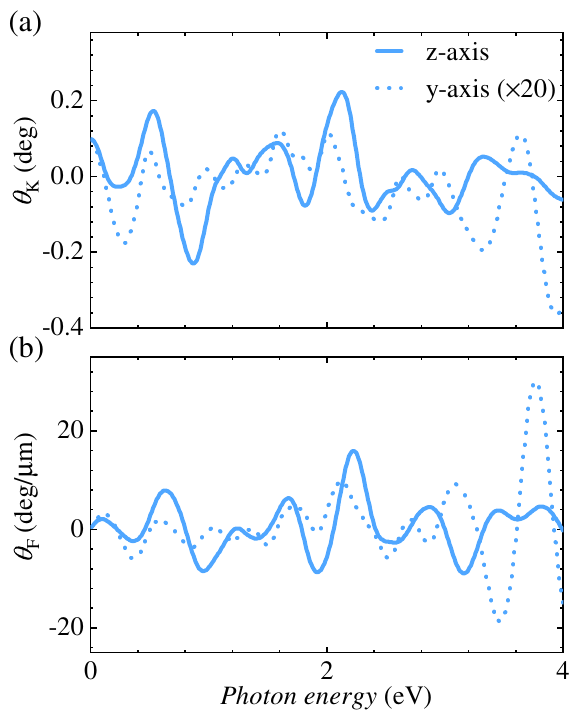}
	\caption{(Color online) The calculated magneto-optical Kerr (a) and  Faraday (b) rotation angles for BL antiferromagnetic F5GT with the out-of-plane (along the $z$-axis) and in-plane (along the $y$-axis) magnetization.  The data of $\theta_\textnormal{K}$ and $\theta_\textnormal{F}$ for the in-plane magnetization are multiplied by a factor of 20.}
	\label{fig:AFM_MO}
\end{figure}

Figure~\ref{fig:AFM_MO} plots the magneto-optical Kerr ($\theta_\textnormal{K}$) and Faraday ($\theta_\textnormal{F}$) rotation angles of BL antiferromagnetic F5GT with the out-of-plane (along the $z$-axis) and in-plane (along the $y$-axis) magnetization.  It can be found that BL antiferromagnetic F5GT also exhibits strong magneto-optical anisotropy as the results of out-of-plane magnetization are much larger than that of in-plane magnetization, similarly to BL ferromagnetic F5GT.  The largest $\theta_\textnormal{K}$ and $\theta_\textnormal{F}$ are -0.23$^\circ$ and 15.94$^\circ/\mu m$ at 0.87 eV and 2.23 eV, respectively.  The $\theta_\textnormal{K}$ of BL antiferromagnetic F5GT is comparable to that of some famous noncollinear and collinear antiferromagnets, such as Mn$_3X$ ($X$ = Rh, Ir, Pt) (0.2 $\sim$ 0.6$^\circ$)~\cite{FengWX2015}, Mn$_3X^\prime$N ($X^\prime$ = Ga, Zn, Ag, or Ni) (0.3 $\sim$ 0.4$^\circ$)~\cite{zhou2019Mn}, RuO$_2$ ($\sim$ 0.62$^\circ$)~\cite{Zhouxiaodong2021}, and CoNb$_3$S$_6$ ($\sim$ 0.2$^\circ$)~\cite{Zhouxiaodong2021}, and is larger than that of noncollinear antiferromagnets Mn$_3Y$ ($Y$ = Ge, Sn) (0.008 $\sim$ 0.02 $^\circ$)~\cite{Higo2018,dWumingxing2020}.  The strong MOE in BL antiferromagnetic F5GT suggests a new material platform for antiferromagnetic magneto-optical devices.

\begin{figure*}
	\includegraphics[width=2\columnwidth]{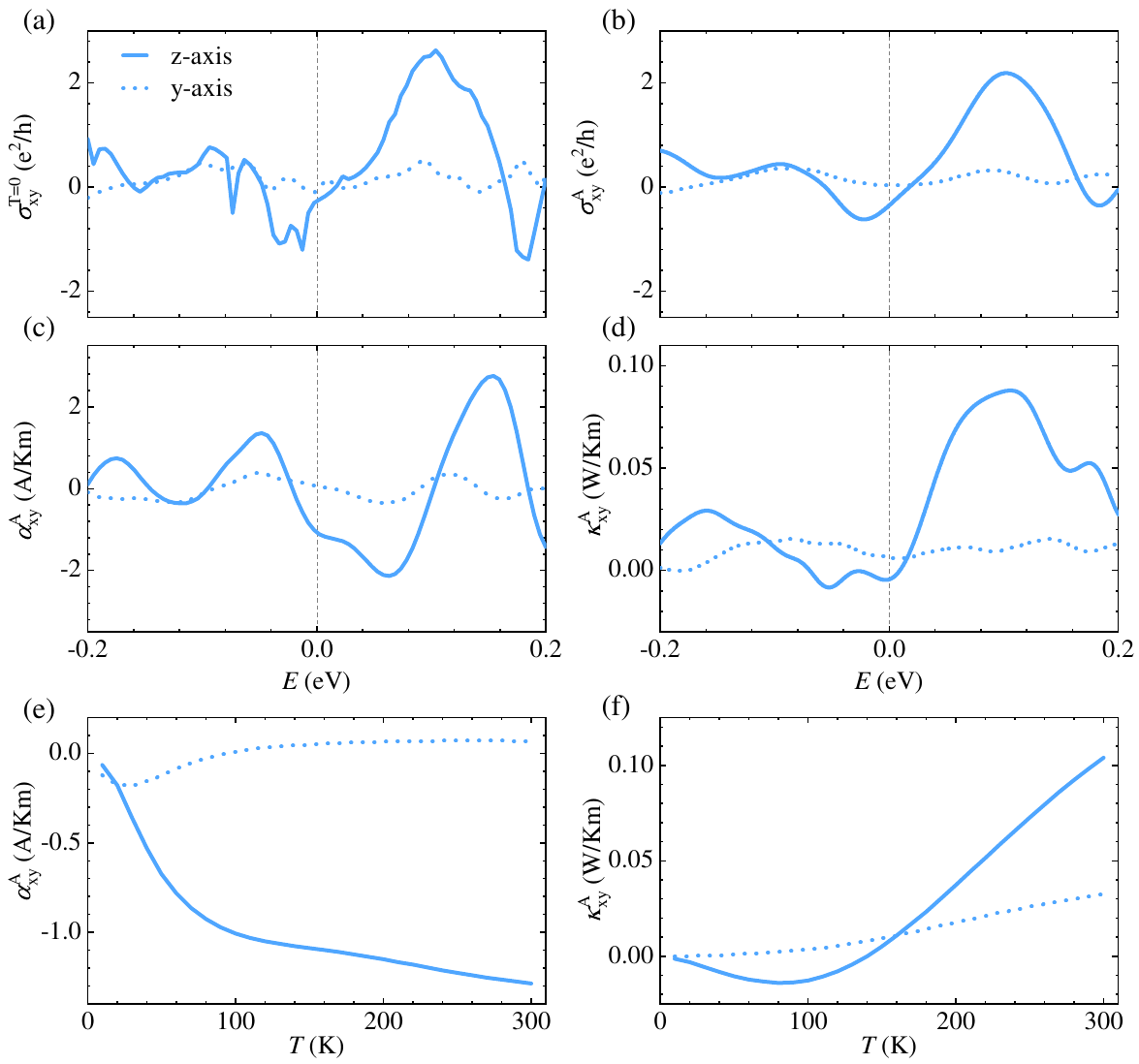}
	\caption{(Color online) The calculated anomalous Hall conductivities at 0 K (a) and 130 K (b), anomalous Nernst conductivity at 130 K (c), and anomalous thermal Hall conductivity at 130 K (d) for BL antiferromagnetic F5GT with the out-of-plane (along the $z$-axis) and in-plane (along the $y$-axis) magnetization as a function of the Fermi energy.  (e) and (f) are the anomalous Nernst and thermal Hall conductivities as a function of temperature calculated at the true Fermi energy.}
	\label{fig:AFM_AHE}
\end{figure*}

Finally, let us turn our attention to the transverse electronic, thermoelectric, and thermal transport properties of BL antiferromagnetic F5GT, as shown in Fig.~\ref{fig:AFM_AHE}.  Like the magneto-optical Kerr and Faraday spectra, the AHC, ANC, and ATHC appeared in the out-of-plane magnetization are much stronger than that in the in-plane magnetization.  For the out-of-plane magnetization, the zero-temperature AHC is -0.28 $e^2/h$ at the $E_F$ and can be significantly enhanced by electron doping, such as 2.63 $e^2/h$ at $E_F+0.10$ eV [Fig.~\ref{fig:AFM_AHE}(a)].  When the temperature increases, the AHC tends to be weaker [Fig.~\ref{fig:AFM_AHE}(b)].  For example, the AHC calculated at 130 K is -0.35 $e^2/h$ at the $E_F$ and the largest value is 2.19 $e^2/h$ at $E_F+0.10$ eV.  The largest zero-temperature AHC (2.63 $e^2/h$ $\approx$ 518 S/cm) is larger than that of some famous noncollinear antiferromagnets, e.g., Mn$_3Y$ ($Y$ = Ge, Ga, Sn) (100 $\sim$ 300 S/cm)~\cite{Guo_anti2017}, and Mn$_3X^\prime$N ($X^\prime$ = Ga, Zn, Ag, or Ni) (-359 $\sim$ 344 S/cm)~\cite{zhou2019Mn}.

Figures~\ref{fig:AFM_AHE}(c) and~\ref{fig:AFM_AHE}(d) depict the ANC and ATHC calculated at 130 K as a function of energy, respectively. The ANC and ATHC are -1.05 A/Km and -0.004 W/Km at the $E_F$, respectively.  Through doping electrons, the ANC can reach up to -2.14 A/Km and 2.76 A/Km at 0.06 eV and 0.15 eV, respectively, which are larger than that of Mn$_3X$ ($X$ = Sn, Ge) (-0.89 $\sim$ -0.54 A/Km)~\cite{Guo_anti2017} and bcc Fe (1.8 A/Km)~\cite{Lixk2017}.  The ATHC reaches up to about 0.10 W/Km at 0.10 eV, which is comparable to that of BL ferromagnetic F5GT and is slightly larger than some famous noncollinear antiferromagnets, such as Mn$_3$Sn (0.04 W/Km)~\cite{Lixk2017} and Mn$_3$Ge (0.015 W/Km)~\cite{XuLC2020}.  Since the N\'eel temperature of BL antiferromagnetic F5GT is unknown yet, we calculate the ANC and ATHC at the $E_F$ as a function of temperature, as shown in Figs.~\ref{fig:AFM_AHE}(e) and~\ref{fig:AFM_AHE}(f).  The dependence of the ANC and ATHC on temperature in the range of 10 $\sim$ 300 K is not much obvious for the in-plane magnetization.  On the other hand, for the out-of-plane magnetization, the temperature has a great influence on the ANC and ATHC.  For example, the ANC and ATHC increase from almost zero at 10 K to -1.28 A/Km and 0.10 W/Km at 300 K, respectively.  Our results suggest that BL antiferromagnetic F5GT is an ideal material platform for the applications of antiferromagnetic spin caloritronics and thermal devices instead of usual ferromagnets.

\section{Summary}

In conclusion, utilizing the first-principles calculations together with group theory analysis, we systematically investigated the MAE, MOE, AHE, ANE, and ATHE in ML and BL Fe$_n$GeTe$_2$ ($n$ = 3, 4, 5).  The calculated MAE indicates that ML and BL F3GT prefer to the out-of-plane magnetization, while the ML and BL F4GT and F5GT are in favor of  the in-plane magnetization.  Moreover, the MAE shows the magnetic isotropy in the $xy$ plane and as the number of layers increases the out-of-plane magnetization enhances.  Additionally, we also confirmed the magnetic ground states of BL Fe$_n$GeTe$_2$.  Except that BL F4GT is energetically stable in its ferromagnetic state, both BL F3GT and BL F5GT have lower energies in the antiferromagnetic states.  We have determined the symmetry requirements of the MOE, AHE, ANE, and ATHE via magnetic group theory analysis.  In short, these closely related physical phenomena can arise in ML and BL ferromagnetic F3GT, F4GT, and F5GT with the $z$-axis magnetization, in BL ferromagnetic F3GT with the $y$-axis magnetization, in ML and BL ferromagnetic F4GT and F5GT with the $y$-axis magnetization, and more intriguingly in BL antiferromagnetic F5GT with the $y$- or $z$-axis magnetization.  The results of group theory analysis are fully consistent with our first-principles calculations.

The MOE, AHE, ANE, and ATHE show the strong anisotropy between the in-plane and out-of-plane magnetization.  The largest magneto-optical Kerr and Faraday rotation angles are found in the ferromagnetic states with the out-of-plane magnetization. The Kerr angles of 1.19$^\circ$ (2.07$^\circ$), 1.14$^\circ$ (1.90$^\circ$), and 0.74$^\circ$ (1.44$^\circ$) are calculated in ML (BL) F3GT, F4GT, and F5GT, respectively; the Faraday angles of 222.66$^\circ/\mu m$ (147.73$^\circ/\mu m$), 109.12$^\circ/\mu m$ (82.95$^\circ/\mu m$), and 50.28$^\circ/\mu m$ (54.07$^\circ/\mu m$) are calculated in ML (BL) F3GT, F4GT, and F5GT, respectively.  Furthermore, the calculated zero-temperature AHC at the $E_F$ are 0.73 (1.57) $e^2/h$, 1.70 (2.54) $e^2/h$, and 0.46 (1.50) $e^2/h$ in ML (BL) ferromagnetic F3GT, F4GT, and F5GT with the out-of-plane magnetization.  By doping electrons or holes, the zero-temperature AHC of BL ferromagnetic F3GT, F4GT, and F5GT can be significantly enhanced to -2.16 $e^2/h$, 4.48 $e^2/h$, and 3.46 $e^2/h$, respectively.  According to the generalized Landauer-B\"uttiker formalism, the temperature-dependent AHC, ANC, and ATHC can be obtained from zero-temperature AHC.  The overall trend of  finite-dependent AHC as a function of the Fermi energy is very similar to that of zero-temperature AHC, but the magnitude of AHC reduces markedly with the increasing of temperature.  At the temperature of 130 K, the largest ANC can reach up to 1.85 (-2.52) A/Km, -1.85 (-3.31) A/Km, and 2.37 (1.68) A/Km in ML (BL) F3GT, F4GT, and F5GT by appropriate electron or hole doping, respectively.  The largest ATHC in ML (BL) F3GT, F4GT, and F5GT are calculated to be 0.16 (0.05) W/Km, 0.14 (0.22) W/Km, and 0.19 (0.16) W/Km at 130 K, respectively.  The noteworthy MOE, AHE, ANE and ATHE discovered in ML and BL ferromagnetic Fe$_n$GeTe$_2$ ($n$ = 3, 4, 5) are comparable to or even larger than that of many familiar ferromagnets, suggesting that 2D Fe$_n$GeTe$_2$ have potential applications in magneto-optical devices, spintronics, spin caloritronics, and etc.

Another interesting finding in our work is that the MOE, AHE, ANE, and ATHE surprisingly appear in BL collinear antiferromagnetic F5GT, which is much rare in 2D magnets.  The first-principles results show that the MOE, AHE, ANE and ATHE in BL collinear antiferromagnetic F5GT are considerably large.  For the out-of-plane magnetization, the largest Kerr angle, Faraday angle, zero-temperature AHC, ANC, and ATHC are -0.23$^\circ$, 15.94$^\circ/\mu m$, -0.28 $e^2/h$, -1.05 A/Km (130 K), and -0.004 W/Km (130 K), respectively.  By doping electrons or holes, the AHC, ANC, and ATHC can increase up to 2.63 $e^2/h$, 2.76 A/Km (130 K) and 0.1 W/Km (130 K), respectively.  On the other hand, by increasing temperature to 300 K, the ANC and ATHC can reach up to -1.28 A/Km and 0.10 W/Km at the $E_F$.  These results are comparable to or even larger than that of some famous noncollinear and collinear antiferromagnets, indicating the BL collinear antiferromagnetic F5GT is a good material for multi-functional device applications based on antiferromagnetism instead of usual ferromagnetism.

\begin{acknowledgments}
The authors thank Xiaoping Li, Jin Cao, Chaoxi Cui, Shifeng Qian, Jun Sung Kim, Xiaobo Lu, and Li Yang for their helpful discussion.  This work is supported by the National Natural Science Foundation of China (Grant Nos. 11874085, 11734003, and 12061131002), the Sino-German Mobility Programme (Grant No. M-0142), and the National Key R\&D Program of China (Grant No. 2020YFA0308800).
\end{acknowledgments}
	
\bibliography{references}

\end{document}